\documentclass[%
aps,prl,twocolumn,showpacs,preprintnumbers,amsmath,amssymb,nofootinbib]%
{revtex4-1}
\usepackage{graphicx}
\usepackage{dcolumn}
\usepackage{bm}
\usepackage{slashed}
\usepackage{amsmath}
\usepackage{mathrsfs}
\newcommand{\disp}{\displaystyle}
\newcommand{\ba}{\begin{array}}
\newcommand{\ea}{\end{array}}
\newcommand\gr[1]{\mathrm{#1}}    
\newcommand\muI{\mu_{\mathrm{I}}}
\newcommand\Nc{N_{\mathrm{c}}}
\newcommand\pic{\bm{\pi}_{\mathrm{c}}}
\newcommand{\ie}{{i.e.}}
\newcommand{\beq}{\begin{equation}}
\newcommand{\eeq}{\end{equation}}
\newcommand{\bea}{\begin{eqnarray}}
\newcommand{\eea}{\end{eqnarray}}
\DeclareMathOperator{\diag}{diag}
\begin{document}
\title{%
Splitting of the chiral critical point and realization 
of solitonic pion condensate driven by isospin density
}%
\author{Yuhei Iwata}
\email[E-mail:~]{j1211604@ed.tus.ac.jp}
\author{Hiroaki Abuki}%
\email[E-mail:~]{h.abuki@rs.tus.ac.jp}
 \author{Katsuhiko Suzuki}
\email[E-mail:~]{katsu\_s@rs.kagu.tus.ac.jp}
\affiliation{%
Department of Physics, Tokyo University of Science, Tokyo 162-8601, Japan}%
\date{\today}
\begin{abstract}
We study the influence of the isospin asymmetry on the phase structure
 of strongly interacting quark matter near the tricritical point (TCP)
 using a generalized Ginzburg-Landau approach.
The effect has proven to be so drastic, not only bringing about the
 shift of the location of TCP, but resulting in a rich fine structure at
 the vicinity of TCP. In particular, we find that an arbitrary small
 perturbation due to isospin density lifts the degeneracy of TCP making
 it split into four independent multicritical points.
Accordingly, the homogeneous pion condensate and its solitonic
 counterpart come to occupy large domains in the Ginzburg-Landau
 coupling space.
\end{abstract}
\maketitle

\emph{Introduction.}---%
Quantum chromodynamics (QCD) at finite temperature and/or finite density
is expected to exhibit a rich phase structure, and has been the subject
of extensive theoretical and experimental studies. In particular,
several approaches to QCD with two light flavors suggest the existence
of a critical end point (CEP) where the first order chiral phase
transition turns into a crossover \cite{Asakawa:1989bq}; the CEP
continues to a tricritical point (TCP) in the chiral limit where three
lines of second-order phase transition meet. Despite many efforts based
on the first principle calculations \cite{Fukushima:2010bq}, the precise
location of CEP in the phase diagram is still controversial.

Recently based on a Ginzburg-Landau (GL) approach \cite{Nickel2009}, and
also in other effective models \cite{Nickel:2009wj}, it was shown that
such TCP, if any, is actually replaced by a Lifshitz point where three
different forms of matter meet; \ie, a phase with broken chiral
symmetry, a symmetric (Wigner) phase, and an inhomogeneous phase
characterised by an additional translational symmetry breaking.
Such inhomogeneous chiral condensates can be viewed as a microscale {\em
ordered} phase separation \cite{Abuki:2011pf,Fukushima:2012mz}.

The effect of a finite quark mass on TCP is rather simple; just to turn
it into a CEP. Our focus here is the other important ingredient in
realistic systems, \ie, the effect of an isospin asymmetry on TCP. Such
a flavor symmetry breaking can be caused by a neutrality constraint
which should be imposed in any bulk system to prevent the energy density
from diverging. The isospin asymmetry effect on the thermodynamics is
also important for the physics of heavy ion collisions with neutron rich
nuclei at the Fermi energies, as it affects the isospin distillation in
the nuclear liquid-gas phase transition \cite{Chen:2004si}.

The isospin asymmetry effects were well studied in the physics of color
superconductivity at high density, and proved to lead to a rich variety
of phases \cite{Abuki:2004zk}. On the other hand, while there are a few
model-based studies in the context of TCP \cite{Klein:2003fy}, there
has been to present, to our knowledge, no systematic analysis based on
the GL approach. Here we present for the first time GL analyses of the
effects on phases at the vicinity of TCP.

By comparison with other approaches, our GL framework has an advantage
that it can give model independent predictions near TCP. Based on it, we
take into consideration the charged pion condensate which may become
favored by the inclusion of isospin chemical potential
\cite{Son:2000xc}. Moreover we incorporate the possibility of
inhomogeneous condensate such as a chiral spiral \cite{Nakano:2004cd} in
which the chiral condensate is entangled with the pion condensate. Since
we are interested in the response of TCP and its neighborhood against
the isospin chemical potential ($\muI$), our strategy is to take $\muI$
as a perturbative parameter and expand the GL functional with respect to
it. As a result we have many GL couplings, but it is possible to
derive universal relations among them; this will be performed based on
the assumption that the quark loops are dominant near the TCP. This is
justified if it is located at large fugacity region $e^{\mu/T}\agt
1$. The analyses for the opposite case will be reported elsewhere.

Our findings are; i) an arbitrary small perturbation due to isospin
density not only brings about the shift of TCP but makes it split into
four independent critical points, and ii) both the homogeneous and
inhomogeneous pion condensates come to occupy large domains in the GL
coupling space. Although it is beyond the scope of this Letter, these
pion condensates near TCP may smoothly continue to pion condensates in
nuclear matter at zero temperature \cite{Migdal:1990vm}, or to those
with some spatial structures \cite{Takatsuka:1978ku}; such meson
condensates are of a renewed interest since their analogs may have a
chance to be realized in ultracold atomic gases as was recently reported
\cite{Maeda:2012gw}. The fine phase structure near TCP may also have
some phenomenological impacts on the physics of heavy ion collisions,
and on the proto-neutron star cooling via the neutrino diffusion.

\emph{Generalized Ginzburg-Landau approach at finite $\muI$.}---%
Let us start with writing the most general GL potential for the chiral
four vector $\phi({\bf x}) = (\sigma({\bf x}), \bm{\pi}({\bf x}))$ with
$\sigma({\bf x})\sim -\langle \bar\psi\psi\rangle$, $\bm{\pi}({\bf
x})\sim-\langle \bar\psi i\gamma^5\bm{\tau}\psi\rangle$. We retain up to
the sixth order in the order parameter and its spatial derivative so as
to allow for a minimal description of TCP. Then the GL functional can be
decomposed into three parts:
\beq
\Omega_{\rm GL}[\sigma({\textbf x}), \bm{\pi}({\textbf x})]=\omega_0%
+ \delta\omega_\mathrm{M} + \delta\omega_\mathrm{I},
\notag
\eeq
with $\omega_0$ being the chiral $\gr{SU(2)_{R}}\times\gr{SU(2)_{L}}$
invariant part, and $\delta\omega_{\mathrm{M}}$
($\delta\omega_{\mathrm{I}}$) being the feedback from current quark mass
(isospin density). The form of each part is rather stringently
constrained by symmetry as we describe below. First, $\omega_0$ can be
set as
\bea
\omega_0 &=& \frac{\alpha_2}{2}\phi^2 + \frac{\alpha_4}{4}(\phi^2)^2 
+\frac{\alpha_{4b}}{4} (\nabla\phi)^2 + \frac{\alpha_6}{6}(\phi^2)^3
\notag\\
&&+\frac{\alpha_{6b}}{6}(\phi, \nabla\phi)^2 +
\frac{\alpha_{6c}}{6}[\phi^2(\nabla\phi)^2-(\phi, \nabla\phi)^2]\notag\\
&&
+\frac{\alpha_{6d}}{6}(\Delta\phi)^2,\notag
\eea
where $(\phi,\nabla\phi)$ denotes the inner
product:~$\sigma\nabla\sigma+\bm{\pi}\cdot\nabla\bm{\pi}$. Second,
$\delta\omega_\mathrm{M} = -h\sigma$ breaks the chiral symmetry
explicitly: $\gr{SU(2)_L} \times\gr{SU(2)_R}\to\gr{SU(2)_{V}}$.
The GL coupling $h$ is proportional to the quark mass $m$ for light
flavors. Lastly, the term $\delta\omega_\mathrm{I}$ represents the
response to isospin density, which is our main focus here,
\beq
\ba{rcl}
\delta\omega_\mathrm{I} &=&
\disp\frac{\beta_2}{2}\pic^2+\frac{\beta_4}{4}\pic^4%
+\frac{\beta_{4b}}{4}(\phi^2-\bm{\pi}_c^2)\pic^2%
+\frac{\beta_{4c}}{4}(\nabla\pic)^2\,,
\ea
\notag
\eeq
where $\bm{\pi}_c=(\pi_1,\pi_2)$ is the charged pion doublet. This term
breaks $\gr{SU(2)_L}\times\gr{SU(2)_R}$ down to
$\gr{U(1)_{I_3;V}}\times\gr{U(1)_{I_3;A}}$. Negative
$\beta_2,\,\beta_4,\,\beta_{4b}$ favor the homogeneous pion condensate
$|\pic|\ne0$ while negative $\beta_{4c}$ does inhomogeneous one. 
It is safe to neglect the quartic terms $\beta_4,\,\beta_{4b},\beta_{4c}$
as long as we are only interested in the extreme vicinity of TCP where
both $|\pic|,\,|\phi|$ are much smaller than the order of $\muI$. We
here retain them because we are interested in the structure of phases
where the order parameters become comparable with $\muI$. To summarise
the symmetry structure, $\gr{SU(2)_L}\times\gr{SU(2)_R}$ is broken to
$\gr{SU(2)_V}$ due to $\delta\omega_\mathrm{M}$, and then further down
to $\gr{U(1)_{I_3;V}}$ via $\delta\omega_\mathrm{I}$. The residual
$\gr{U(1)_{I_3;V}}$ may be broken spontaneously via the formation of
charged pion condensate.

\begin{figure}[t]
\begin{center}
\includegraphics[width=0.95\linewidth]{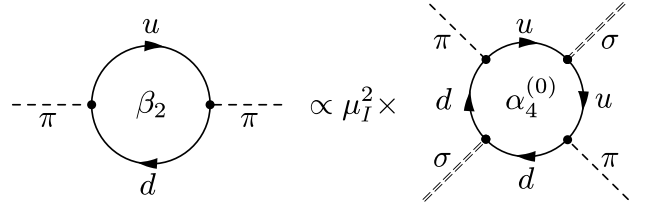}
\caption{The Feynman diagram for $\beta_2$, which is the leading
 driving force to the pion condensate when $\alpha_4^{(0)}>0$.}
\label{fig:FD}
\end{center}
\end{figure}

Ignoring the quark mass term $\delta\omega_\mathrm{M}$, we have eleven
couplings $\{\alpha_2,\alpha_4,\alpha_{4b},\alpha_6,\alpha_{6,b},%
\alpha_{6,c},\alpha_{6,d},\beta_2,\beta_4,\beta_{4b},\beta_{4c}\}$.
We can reduce the number of couplings significantly using the expansion
about $\muI=0$ as we demonstrate below. In order to make this, we assume
that at the vicinity of TCP quark loops are dominant to gluonic ones,
which may be justified provided that TCP is located at large fugacity
region. The feedback of quark loops to the energy is
\beq
\Delta\Omega=-\frac{T\Nc}{V}
\sum_{n=2}\frac{1}{n}\mathrm{Tr}\left(S_0 \Sigma({\textbf x})\right)^n,
\notag
\eeq
where $V$ denotes the volume of the unit cell of periodic condensate,
and $\gr{Tr}$ should be taken over the Dirac, flavor and functional
indices. $S_0 = \diag{(S_u, S_d)}$ is the bare quark propagator, and
$\Sigma({\textbf x}) =\sigma({\textbf x}){\bf{1}} + i\gamma^5\pi({\bf
x})\tau^1$ is the self-energy. Since the potential is symmetric under
$\gr{U(1)_{I_3;V}}\times\gr{U(1)_{I_3;A}}$ that leaves
$\sigma^2+\pi_0^2$ and $\pic^2$ independently invariant, we set
$\pi_2=\pi_0=0$ without any loss of generality to derive the GL
couplings. The propagator in the momentum space is
$S(p)=\mathop{\mathrm{diag}}(\slashed p_u^{-1}, \slashed p_d^{-1})$ with
$p_{f}^\mu=(i\omega_m+\mu_{f},{\textbf p})$ where the subscript $f$ refers to
the flavor index for $u$ and $d$ quarks, and $\omega_m$ is the fermionic
Matsubara frequency. We do not show all the results, but for example
$\beta_2$ and $\alpha_4$ have the expressions;
$\beta_2=4T\Nc\sum_{m,{\textbf p}}\frac{(p_u-p_d)^2}{p_u^2p_d^2}$,
$\alpha_4=4T\Nc\sum_{m,{\textbf p}}\frac{p_u^4+p_d^4}{p_u^4p_d^4}$. Let us
start with the GL couplings $\alpha$'s which enter in $\omega_0$. From
the expressions for $\alpha$'s, we can derive the universal relations
$\alpha_{4,b} = \alpha_4$,
$(\alpha_{6,b},\alpha_{6,c},\alpha_{6,d})=(5,3,1/2)\alpha_6$ \cite{Nickel2009}.
Although $\omega_0$ is the chirally symmetric part, the couplings
themselves can have dependence on $\mu_I^2$, whose expansion should
have the following general structure:
\bea
\begin{pmatrix}
\alpha_2\\ 
\alpha_4\\
\alpha_6
\end{pmatrix}
=
\begin{pmatrix}
  1 & a\muI^2 & \mathcal{O}(\muI^4)\\
  0 & 1       &  b\muI^2\\
  0 & 0       & 1 \\
\end{pmatrix}%
\begin{pmatrix}%
\alpha_2^{(0)}\\
\alpha_4^{(0)}\\
\alpha_6^{(0)}%
\end{pmatrix},
\notag
\eea
where $a$, $b$ are just some constants and the subscript $(0)$ denotes
the quantity in the absence of isospin
density. Explicit computation leads $a=0$, $b=1$.
Similarly 
the expansion of $\delta\omega_\mathrm{I}$
is cast into the form:
\bea
\begin{pmatrix}
\beta_2\\
\beta_4\,[\mbox{or }\beta_{4b(c)}]
\end{pmatrix}
=\begin{pmatrix}
 c\muI^2 & \mathcal{O}(\muI^4)  \\
  0  & d\muI^2\,[d_{b(c)}\muI^2]\\
\end{pmatrix}%
\begin{pmatrix}
\alpha_4^{(0)}\\
\alpha_6^{(0)}
\end{pmatrix},%
\notag
\eea
and we found $c=-1/2$, $d=d_{b(c)}=-1$. When $\alpha_4^{0}>0$,
$\beta_2=-\muI^2\alpha_4^{(0)}/2$ acts as the main driving force to the
formation of homogeneous pion condensate, whose diagrammatic expression
is shown in Fig.~\ref{fig:FD}. Since we can use $\alpha_6^{(0)}(>0)$ to
set the energy scale, we are left with only three parameters
$\{\alpha_2^{(0)},\alpha_4^{(0)},\muI^2\}$. We study the phase diagram
in the space of $\{\alpha_2^{(0)},\alpha_4^{(0)}\}$ and how it is
affected by $\muI^2$. It has in principle a unique map onto the QCD
phase diagram at the vicinity of the critical point where the chiral
condensate has a size less than or of order $\muI$.

Just to avoid notational confusion, we suppress the subscript $(0)$
for $\alpha$'s in the following. Moreover we scale every quantity with
energy dimension in the unit $[\alpha_6^{-1/2}]$. In particular we have
$\alpha_6=1$. Using the scaling argument as in \cite{Abuki:2011pf}, it
turns out that any critical condition is of the form
$f(\alpha_2/\alpha_4^2,\muI^2/|\alpha_4|)=0$ where $f$ is some function
having two arguments. From this, we see that even when $\muI$ is
magnified as $\lambda\muI$ with $\lambda$ being an arbitrary real
number, the phase diagram stays the same once we redefine
$(\alpha_2,\alpha_4)$ by $(\lambda^4\alpha_2,\lambda^2\alpha_4)$. In
particular, we conclude that the coordinate of any critical point in
$(\alpha_2,\alpha_4)$-plane, if any, should scale as
$\alpha_2\propto\muI^4$, $\alpha_4\propto\muI^2$.

\begin{figure}[t]
\begin{center}
\includegraphics[width=50mm]{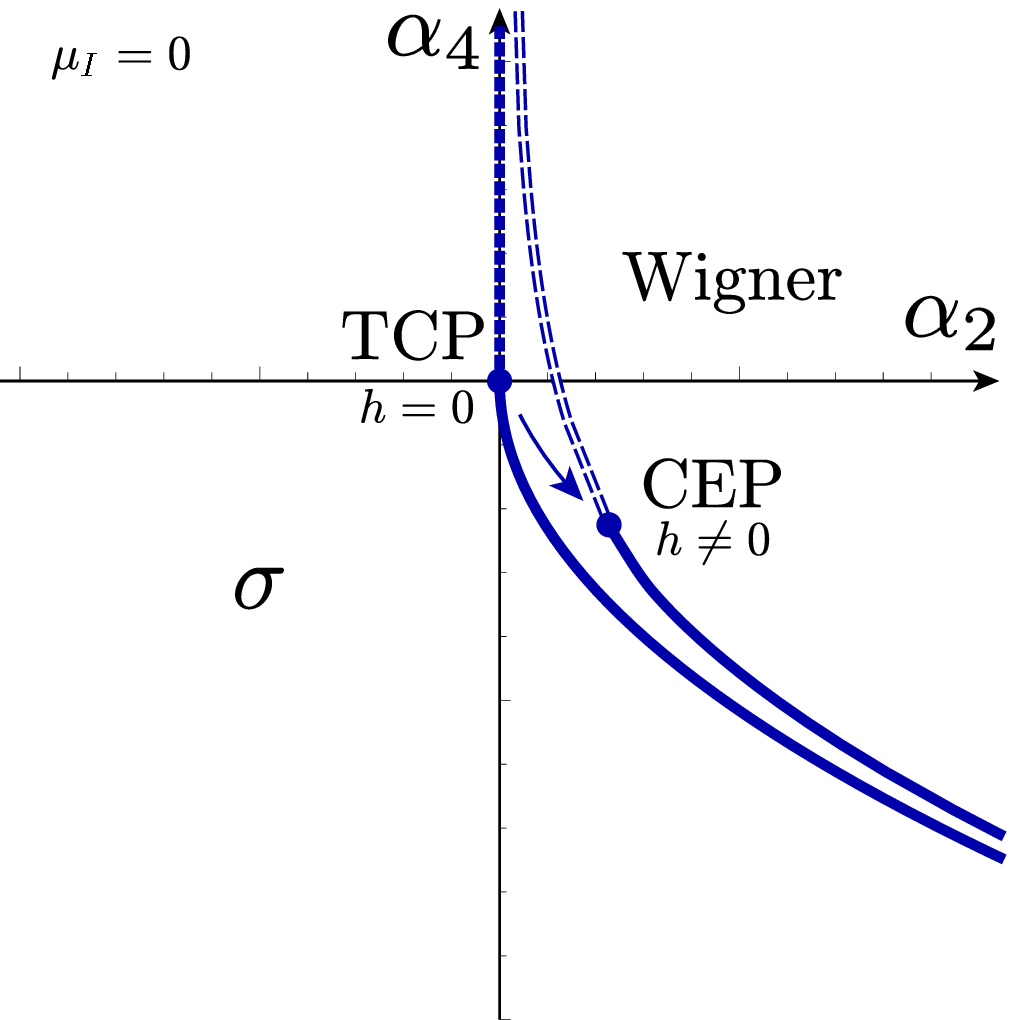}
\includegraphics[width=50mm]{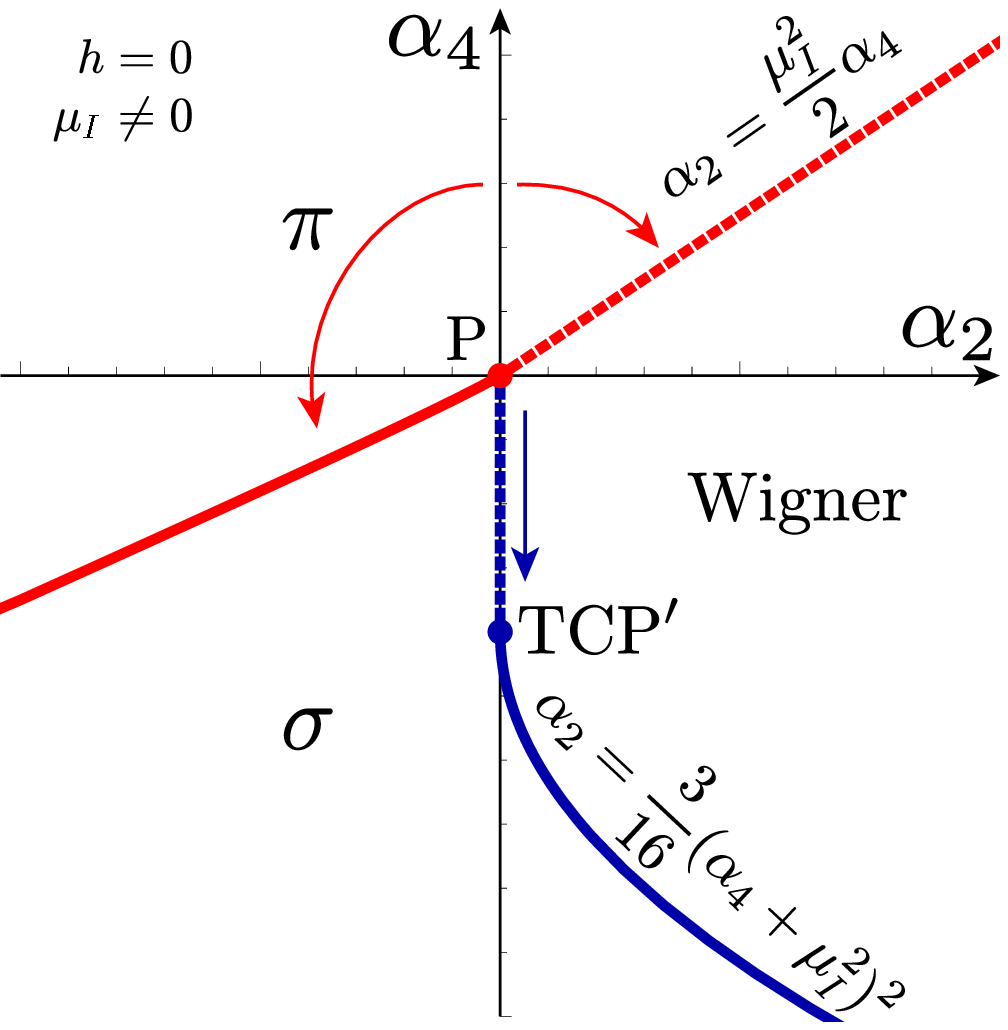}
\caption{(color online). 
Phase structure for homogeneous condensates: Solid (dashed) line
 represents first (second) order transition, while double dashed line
 does the pseudo critical line for crossover. (Upper panel): The effect
 of $h$ in the absence of $\muI$. When $h = 0\,$, TCP is located at the
 origin; we have a second order critical line $\alpha_2=0$ for $\alpha_4
 \ge 0$ and a first order one $\alpha_2 = \frac{3}{16} \alpha_4^2$ for
 $\alpha_4<0$. Nonzero $h$ turns TCP into a CEP whose location is
 shifted to $(2.28h^{4/5}, -2.25h^{2/5})$. (Lower panel): The effect of
 $\muI$ for $h=0$. Asymptotic behaviours of the first order critical
 line separating the $\sigma$ phase from the $\pi$ phase can be derived
 analytically: $\alpha_2\to\frac{3\muI^2}{4}\alpha_4 + \frac{\muI^4}{24}
 +\mathcal{O}(\muI^6)$ as $\alpha_4 \to -\infty$, while $\alpha_2 \to
 \frac{\muI^2}{2}\alpha_4 + \frac{1}{8}\alpha_4^2$ as $\alpha_4\to 0$.
}
\label{fig:fig1}
\end{center}
\end{figure}

\emph{Phase diagram with homogeneous states only.}---%
Let us first discuss the consequence of nonzero $\muI$ to homogeneous
condensates. We show the phase diagram in the lower panel of
Fig.~\ref{fig:fig1}. Just for comparison, we also show in the upper
panel how $\delta\omega_\mathrm{M}$ affects TCP and its neighborhood in
the absence of $\muI$. In the latter case we actually find CEP at
$(2.28h^{4/5},-2.25h^{2/5})$ \cite{Friman:2012gg}. In contrast, we find
a drastic change of phase diagram due to nonzero $\muI$ in the former
case. As expected, we find that pion condensate replaces a major part of
the phase of chiral condensate in particular for $\alpha_4>0$. This is
because $\beta_2=-\muI^2\alpha_4/2<0$ favors the pion condensation. In
fact the second-order phase boundary between the Wigner and pion
condensed phases can be derived from the condition of vanishing
quadratic term:
\beq
0=\left.\frac{\partial^2\Omega_{\rm GL}[0,\pi]}{\partial\pi^2}\right|_{\pi=0}%
=\alpha_2-{\muI^2}\alpha_4/2\,.
\notag
\eeq
Another notable point in the phase diagram is the shift of the location
of TCP from $(0,0)$ to $(0,-\muI^2)$ labeled by TCP${}^\prime$. The
shift itself is consistent with model-based studies \cite{Klein:2003fy},
and the location can be understood by noting
\beq
\Omega_{\rm GL}[\sigma,0]=\frac{\alpha_2}{2}\sigma^2 +
\frac{\alpha_4+\muI^2}{4}\sigma^4 + \frac{1}{6}\sigma^6\,,
\notag
\eeq
where we see that $\alpha_2=0$ together with $\alpha_4+\muI^2=0$ defines
a new TCP. For the region of ($\alpha_2<0$, $\alpha_4<0$), there is a
competition between the chiral and pion condensates, leading to a first
order phase transition between two phases as depicted in the figure. The
coordinate of characteristic points is listed in the
Table~\ref{tab:tab1}.
\begin{table}[b]
    \caption{Location of critical points; CEP, TCP${}^\prime$ and P in
 Fig.~\ref{fig:fig1}, Q, R in  Fig.~\ref{fig:fig2}.
}
\begin{tabular}{|c|c|c|l|}
\hline
    & $\alpha_2$    
    & $\alpha_4$     
    & classification (type) \\ \hline
CEP &
     $\frac{5\mathstrut}{4\mathstrut}\frac{3^{4/5}\mathstrut}{2^{2/5}\mathstrut}h^{4/5}$
    & $-\frac{5\mathstrut}{2^{1/5}3^{3/5}\mathstrut}h^{2/5}$ 
    & critical end point \\ \hline
TCP${}^\prime$ 
    & $0$
    & $-\muI^2$     
    & Lifshitz tricritical point \\
P   & $0$
    & $0$            
    & bicritical point \\
Q   & $3\muI^4/32$ 
    & $-3\muI^2/2$  
    & Lifshitz bicritical point \\
R   & $0.21\muI^4$ 
    & $-2.22\muI^2$ 
    & critical (end) point \\
\hline
\end{tabular}
\label{tab:tab1}
 \end{table}

\begin{figure}[t]
\begin{center}
\includegraphics[width=50mm]{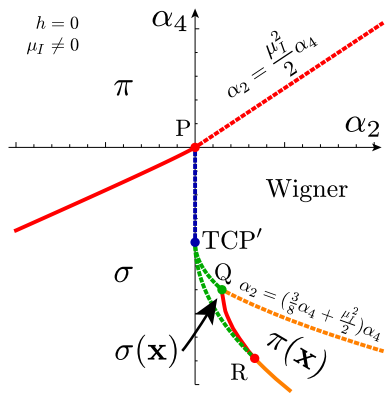}
\caption{(color online). 
Phase structure for $\muI\ne0$.
The second order critical line connecting TCP${}^\prime$ to Q
 is $\alpha_2=\frac{3}{8}(\alpha_4+\muI^2)^2$.
 The second order critical line linking TCP${}^\prime$ to R is
 $\alpha_2 = \frac{5}{36} (\alpha_4 + \muI^2)^2$. The first order
 critical line between the $\sigma(z)$ and $\pi(z)$ phases (a curve
 QR) requires a numerical computation.
}
\label{fig:fig2}
\end{center}
\end{figure}

\emph{Phase diagram with inhomogeneous states included.}---%
We now address the question what is the impact of inclusion of
inhomogeneous phases. We restrict the crystal structure to one
dimensional ones \cite{Abuki:2011pf,Carignano:2012sx}. We analyse three
cases, (i) $\sigma(z)\neq0$, $\pi(z)=0$, (ii) $\sigma(z)=0$,
$\pi(z)\neq0$ and (iii) $\sigma(z)\neq0$, $\pi(z)\neq0$. For the case
(i) or (ii), we can solve analytically the Euler-Lagrange (EL) equation
\cite{Buzdin:1997gg,Nickel2009,Abuki:2011pf}, leading to a solitonic
condensate
\beq 
\sigma(z)\,[\mbox{or } \pi(z)] = \sqrt{\nu} k\, {\mathrm{sn}}{(kz; \nu)}\,,
\notag
\eeq
with $\nu$ denoting the elliptic modulus. In the case (iii), we can not
solve the EL equation analytically. We take here the variational method
instead. The variational state we consider here is one so called chiral
spiral \cite{Nakano:2004cd} where $\sigma$ and $\pi$ are
entangled:~$\sigma(z) = m\cos{(qz)},\,\pi(z) = m\sin{(qz)}$.
This assumption is motivated by the study of the Gross-Neveu model
\cite{Basar:2009fg}. We compute the phase diagram by compering energies
for the three cases. Resulting phase diagram is displayed in
Fig.~\ref{fig:fig2}. We find no window for the chiral spiral, but notice
a new fine structure appearing below TCP${}^\prime$.
TCP${}^\prime$ changes into a Lifshitz point from which a solitonic
chiral condensate ($\sigma({\textbf x})$) expands between the
homogeneous $\sigma$ and Wigner phases; this is indeed expected 
\cite{Nickel2009}. In this case, however, we notice that
$\sigma({\textbf x})$-phase does not continue to the region of
$\alpha_4\ll-\muI^2$; a major part of solitonic $\sigma$ island is taken
over by a solitonic pion condensate $\pi({\textbf x})$. This is caused by
the term $-(|\alpha_4|-\muI^2)(\nabla\sigma)^2$ in the potential where
$\muI^2$ disfavors $\sigma({\textbf x})$. Since the quartic term affects
thermodynamics via the square of its coefficient,
$\sim\alpha_4^2-2\muI^2|\alpha_4|$, the effect overwhelms the other
ones at negative large $\alpha_4$. This makes $\pi({\textbf x})$-phase
replace $\sigma({\textbf x})$-phase eventually. As a consequence there
appear two additional multicritical points denoted by Q and R in the
figure. The precise locations of these points are listed in the
Table.~\ref{tab:tab1}.

Let us briefly discuss the effect of quark mass. Since it breaks
explicitly the chiral symmetry down to $\gr{SU(2)_V}$, the phase
boundary between the Wigner and the $\sigma\ne0$ phase would be smoothen
out. We expect, however, that pionic phases will not be affected so
dramatically since it is characterized by the spontaneous breaking of
$\gr{U(1)_{I_3;V}}$ and so is nothing to do with the explicit breaking
of axial symmetry. The detailed analyses are now under investigation
and will be reported elsewhere.

In conclusion, we performed a systematic GL analysis on the effect of
isospin asymmetry on TCP and phases of its neighborhood. By
incorporating the effect of isospin density perturbatively at the
leading order, we first derived the GL potential which works at the
vicinity of TCP where the order parameters become comparable with
$\muI$. Based on it, we studied how the isospin asymmetry affects the
phase structure. We found that it has several remarkable effects; it
does not only cause a shift of the location of TCP, but also brings
about the development of sizable region for homogeneous and
inhomogeneous pion condensates. This in turn leads to the appearance of
several new multicritical points.

Let us finally make some speculative remarks about the phases found near
TCP. First, although we found a first order phase transition between the
solitonic $\sigma({\textbf x})$ and $\pi({\textbf x})$ phases, it may be
replaced with a fine structure once we determine a suitable functional
form for two condensates solving the coupled EL equation; for instance, a
phase with $\sigma({\textbf x})$ entangled with $\pi({\textbf x})$ may
show up. Second, the pion condensed phases near TCP may continue
smoothly to those discussed in nuclear matter at low temperature
\cite{Migdal:1990vm}. Lastly, multicritical points observed near
TCP${}^\prime$ may have some experimental signatures such as those
discussed for CEP \cite{Stephanov:1998dy}. These clearly deserve further
investigations.

\acknowledgments
Y.I. and H.A. thank J.~Usukura for several useful comments, and
A.~Watanabe for an advice on numerics. A part of numerical calculations
was carried out on SR16000 at YITP in Kyoto University.

\end{document}